\documentclass[3p, twocolumn,sort&compress]{elsarticle}

\journal{Journal of Medical Physics}

\usepackage[english]{babel}
\usepackage{graphicx,tabularx,booktabs,enumitem,multirow}
\usepackage[table]{xcolor}
\usepackage{subcaption}
\usepackage[hyphens]{url}

\usepackage{lineno}
\usepackage{amsfonts,amsmath,amssymb,bm}

\usepackage[linkcolor=black,citecolor=black,filecolor=black]{hyperref}
\hypersetup{pdftitle={},pdfauthor={Lorenzo Mercolli},pdfsubject={},pdfkeywords={}}

\modulolinenumbers[5]

\newcommand{\eq}{ \; = \; }

\begin{document}

\begin{frontmatter}

  \title{Towards Quality Management of Machine Learning Systems for Medical Applications}

  \author[1]{Lorenzo Mercolli\corref{cor1}}  \ead{lorenzo.mercolli@insel.ch}
  \author[1]{Axel Rominger} 
  \author[1]{Kuangyu Shi}

  \cortext[cor1]{Corresponding author}
  \address[1]{Department of Nuclear Medicine, Inselspital \\ Bern University Hospital, University of Bern \\Freiburgstrasse 18, CH-3010 Bern \\ Switzerland }

  \begin{abstract}
    The use of machine learning systems in clinical routine is still hampered by the necessity of a medical device certification and/or by difficulty to implement these systems in a clinic's quality management system. In this context, the key questions for a user are how to ensure reliable model predictions and how to appraise the quality of a model's results on a regular basis.

    In this paper we first review why the common out-of-sample performance metrics are not sufficient for assessing the robustness of model predictions. We discuss some conceptual foundation for a clinical implementation of a machine learning system and argue that both vendors and users should take certain responsibilities, as is already common practice for high-risk medical equipment. Along this line the best practices for dealing with robustness (or absence thereof) of machine learning models are revisited. 

    We propose the methodology from AAPM Task Group 100 report no. 283 \cite{AAPMTG100} as a natural framework for developing a quality management program for a clinical process that encompasses a machine learning system. This is illustrated with an explicit albeit generic example. Our analysis shows how the risk evaluation in this framework can accommodate machine learning systems independently of their robustness evaluation. In particular, we highlight how the degree of interpretability of a machine learning system can be accounted for systematically within the risk evaluation and in the development of a quality management system. 
  \end{abstract}
  
  
\end{frontmatter}

\section{Introduction}

The availability of computational power and of large amounts of data made it possible for artificial intelligence (AI) and machine leading (ML) to become one of the most rapidly developing field of science and technology over the last two decades. The potential of ML in healthcare was quickly recognized and research has been directed first to medical domains where large amounts of more or less standardized data are generated. This includes disciplines that work with image data, such as radiology, nuclear medicine, radiation oncology or pathology (see e.g. Refs.~\cite{Giger2018, Sahiner2019, Shen2020, Maier2019, Madabhushi2016}). The benefits of ML systems in medicine, in particular medical imaging, are manifold and span different areas and modalities, such as e.g. automatic segmentation and contouring \cite{Lustberg2018,Ermics2020}, image reconstruction and quality \cite{Wang2015, Xue2021, Hong2018, Xu2022, Wang2018}, lesion detection and characterization \cite{Zhao2020, Xu2018}, treatment planning \cite{Xue2020, Wang2020,Xue2022}, etc. 
 
While research on ML applications in medicine has grown rapidly, the use of such tools in clinical routine has not yet become standard practice. Like every other software, also ML tools have to fit into a well-defined and potentially rather complex clinical process. Furthermore, the software has to be classified as a medical device and therefore needs to satisfy high standards of robustness and reproducibility. The question of how AI tools can be classified as medical device is an ongoing discussion, as can be seen e.g. in Refs.~\cite{FDA2021,Muehlematter2021,Hwang2019,Benjamens2020}. 

Of course, a risk assessment as well as quality assurance (QA) and quality management (QM) of ML tools play a key role for bringing the potential benefits of AI into clinical routine.  
Without being able to judge and possibly mitigate the potential hazards of an ML system, a clinical implementation is bound to fail. The aim of this paper is to propose a conceptual framework for clinical QM of AI/ML tools and to hint towards the elaboration of robust clinical workflows that include AI. 

Robustness of ML algorithms is not only a challenge in the medical domain. Safety \& Security, financial markets, etc. cannot rely on software, that is overly sensitive to changes in the input. In this paper we therefore start by reviewing the issue of robustness in AI. We construct a simple adversarial example for an image segmentation task to illustrate how small, but carefully constructed perturbations of an input image can cause a failure of a ML model. We will discuss how such adversarial examples can provide a measure of robustness of a model and how they provide an ideal framework for risk assessments under extreme conditions (see also Ref.~\cite{Paschali2018}). Based on this discussion of robustness, we promote a shared responsibility between the software developers/vendors and the users. 

In the second part of this paper, we argue that the AAPM Task Group 100 methodology, described in the report no. 283 \cite{AAPMTG100}, provides a convenient framework to develop a QM system for a clinical workflow that contains AI. The methodology of Ref.~\cite{AAPMTG100} was developed for complex and high-risk clinical processes in radiation oncology. It can be used for e.g.\ proton or ion therapy facilities, which are immensely complex structures with uncountable components that can fail (see e.g.\ Ref.~\cite{Flanz2016}). With a simple example of a generic imaging workflow, we show how the AAPM Task Group 100 methodology \cite{AAPMTG100} can be used to develop a QA program for AI. Furthermore, we discuss how methods from adversarial defenses and interpretable ML can be taken into account in a systematic way. 


\section{Adversarial examples} \label{s:robustness}

It was realized almost a decade ago, that it is actually simple to construct an input for a trained ML model that causes erratic predictions \cite{Szegedy2013} (see also the reviews \cite{Akhtar2018, Biggio2018, Miller2020, Jin2020, Xu2020, Liu2021}). Such an input is called \emph{adversarial example}. The basic idea behind adversarial examples is to design small perturbation to the model input, which are hardly perceivable to humans and which will cause the ML model to make a wrong prediction. As an illustrative example, we shall illustrate these effects in the case of a segmentation task. We will then review how adversarial examples can be used to increase a model's robustness and discuss the best practices.

\subsection{Adversarial examples for image segmentation}\label{s:adv_ex}

\begin{figure*}[tbh]
  \centering
  \subcaptionbox*{\phantomsubcaption}{\includegraphics[width=0.29\textwidth]{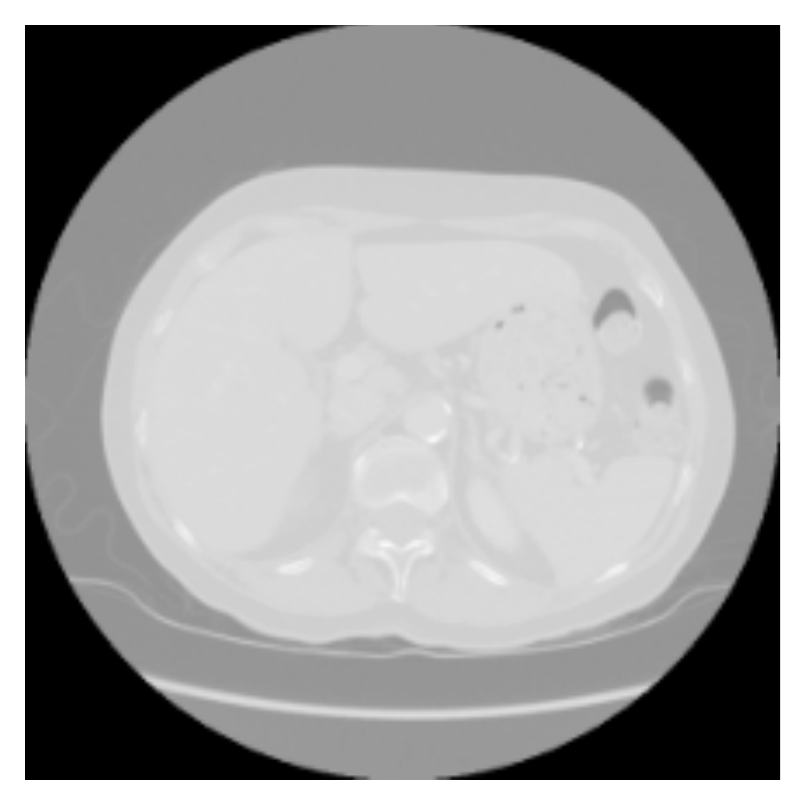} }
  \hfill
  \subcaptionbox*{\phantomsubcaption}{\includegraphics[width=0.29\textwidth]{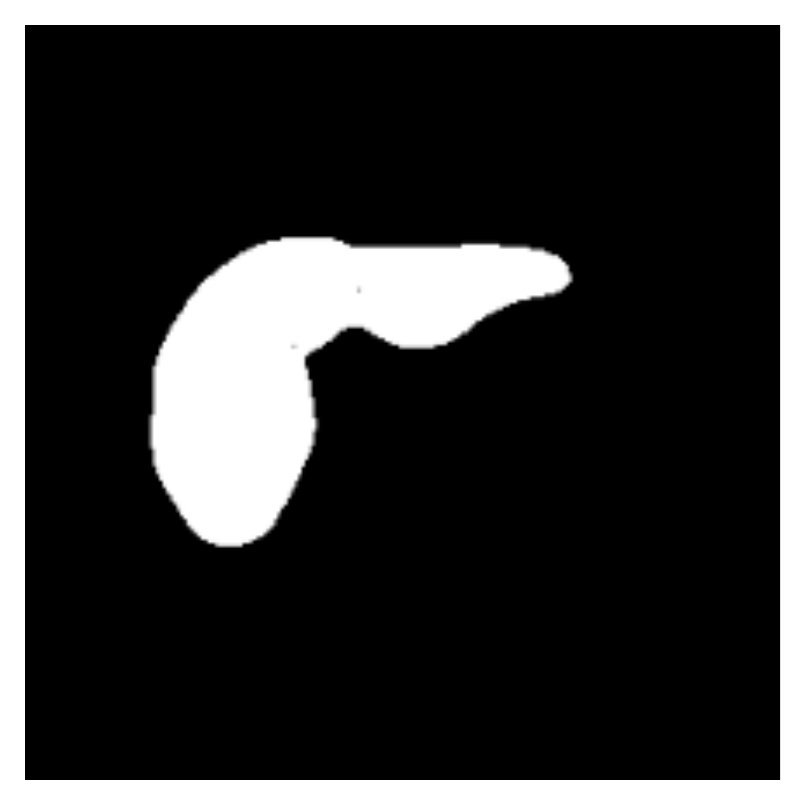} }
  \hfill 
  \subcaptionbox*{\phantomsubcaption}{\includegraphics[width=0.29\textwidth]{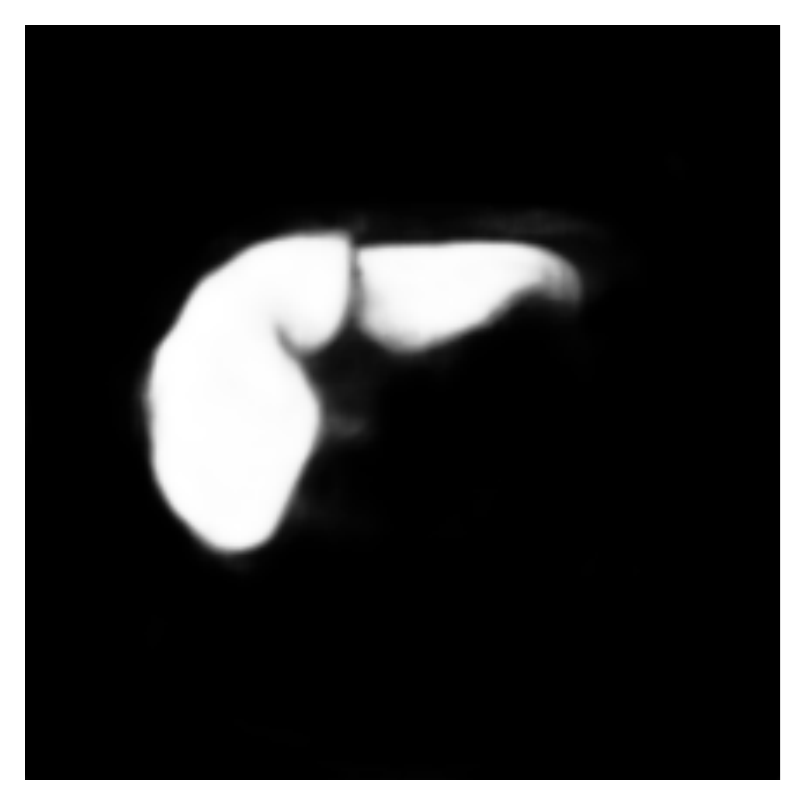} }
  %
  %
  \subcaptionbox*{\phantomsubcaption}{\includegraphics[width=0.29\textwidth]{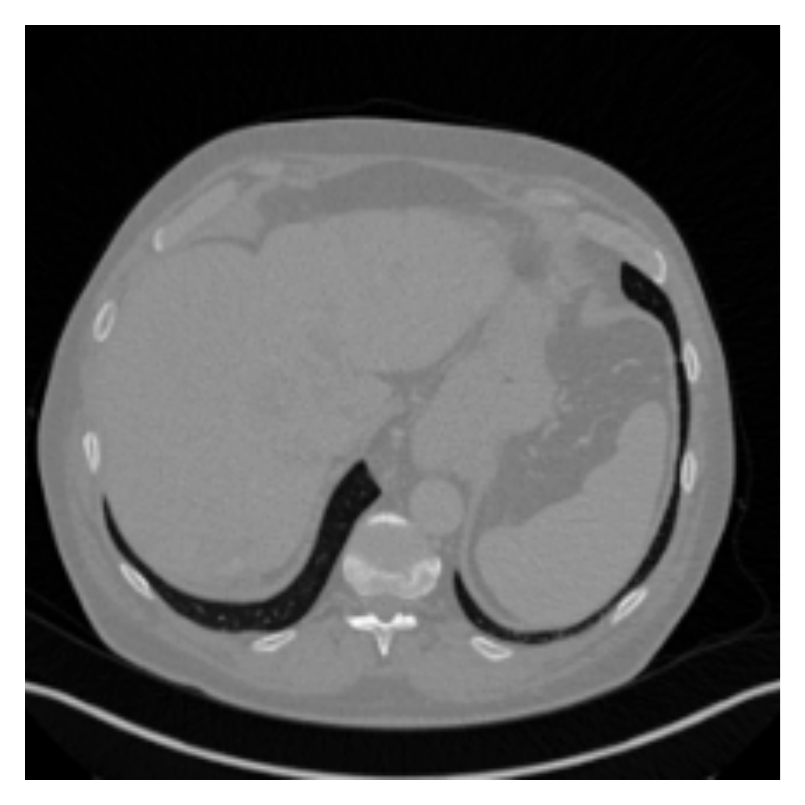} }
  \hfill 
  \subcaptionbox*{\phantomsubcaption}{\includegraphics[width=0.29\textwidth]{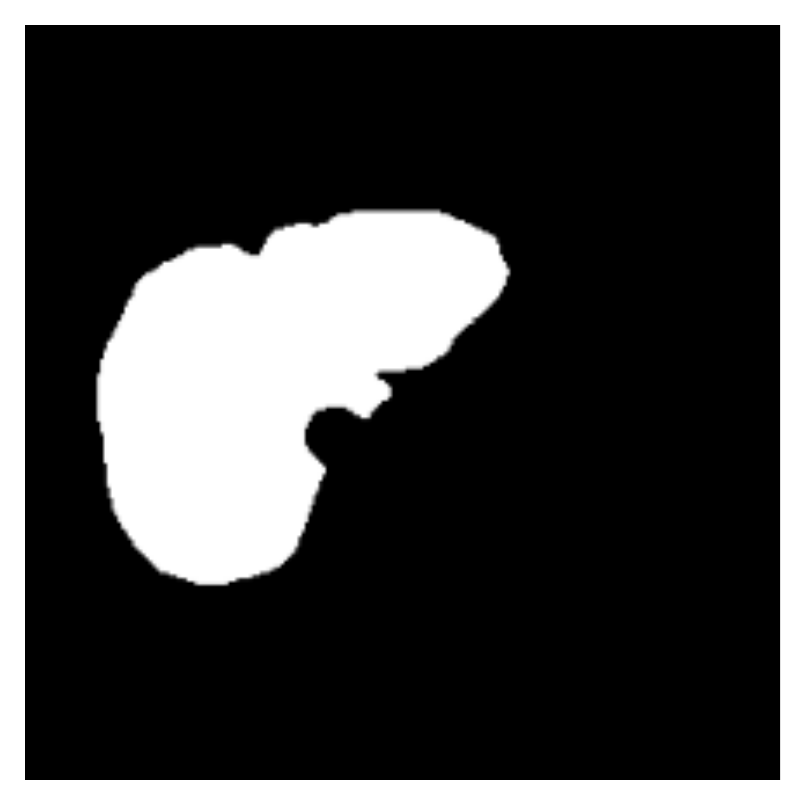} }
  \hfill 
  \subcaptionbox*{\phantomsubcaption}{\includegraphics[width=0.29\textwidth]{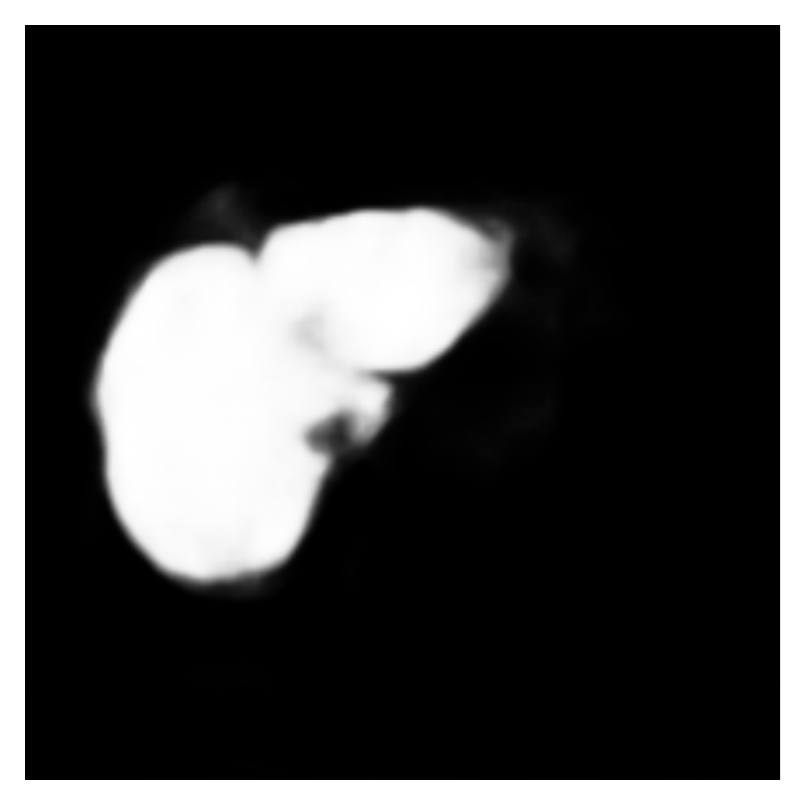} }
  \caption{Two CT images (left column) with the corresponding liver masks (middle column) from the test set are shown together with the model predictions (right column).}\label{f:model_prediction}
\end{figure*}

In medical imaging, a lot of effort has been put into constructing ML models for segmentation tasks (see e.g. Refs.~\cite{Lustberg2018,Ermics2020}). 
For segmentation tasks u-net models \cite{Ronneberger2015} have shown to be very powerful and to perform better than standard convolutional neural networks (CNN). The architecture of the model we use for our example is fairly standard: four down- and four up-sampling blocks with 2D convolutional layers. The model is trained on the liver segmentation data set published in Ref.~\cite{Bilic2019}. This data set contains computer tomography (CT) images of 131 patients with liver tumors (see Fig.~\ref{f:model_prediction} for two example images). The trained model performs well with a focal loss \cite{Lin2017} of $6.29 \cdot 10^{-4}$ on the test set (S{\o}rensen-Dice coefficient \cite{Sorensen1948,Dice1945} of $0.857$ on the test set). In Fig.~\ref{f:model_prediction} we show the model's ability to predict a mask for the liver from a CT image. 

In the literature, there are multiple approaches to constructing adversarial examples. Let us consider the simple \emph{fast gradient sign} method that was put forward in Ref.~\cite{Goodfellow2014}. This is an untargeted attack, meaning that the model should produce any wrong prediction. On the other hand, targeted attacks aim to produce a specific output, i.e.\ force the model to predict a specific target class. Let us denote the loss function of the model prediction $f_\theta(x)$ from an arbitrary input $x$ as $\mathcal{C}(f_\theta(x), y) $. Rather than computing the gradient of $\mathcal{C}(f_\theta(x), y) $ with respect to the model parameters $\theta$, as is done during training, we consider $\nabla_x \mathcal{C}(f_\theta(x), y) $. For a small parameter $\epsilon$, the fast gradient sign adversarial example is simply 

\begin{equation}
  x_{adv} \eq x + \epsilon \cdot  \mathrm{sign} \left( \nabla_x \mathcal{C}(f_\theta(x), y) \right) \;. 
\end{equation}
With the parameter $\epsilon$ we can make the perturbation of the original image $x$ sufficiently small, such that it is hardly recognized by a human. 

\begin{figure*}
  \centering
  \subcaptionbox{\label{f:fs_original}}{\includegraphics[width=0.3\textwidth]{paper_plots/example_image_5.pdf} }
  \hfill 
  \subcaptionbox{\label{f:fs_adv_ex}}{\includegraphics[width=0.3\textwidth]{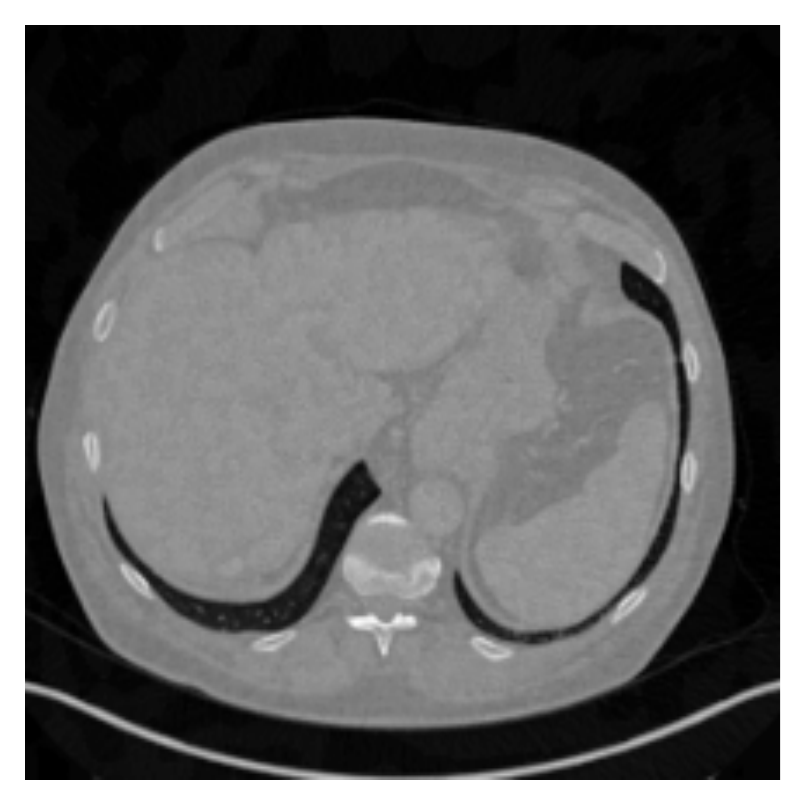} }
  \hfill
  \subcaptionbox{\label{f:fs_adv_ex_pred}}{\includegraphics[width=0.3\textwidth]{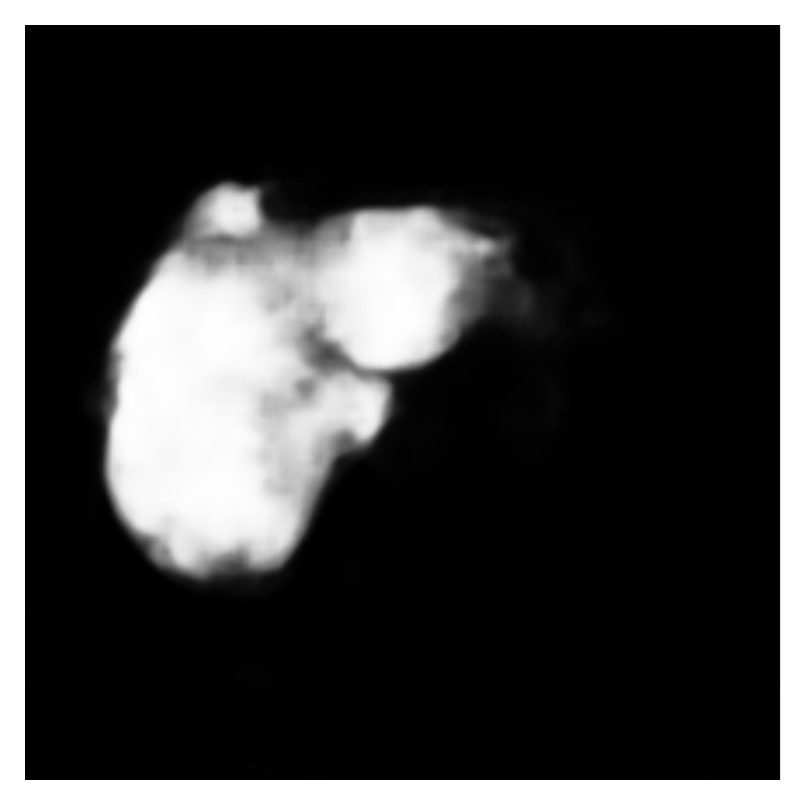} }
  \caption{The original CT image \ref{f:fs_original} together with the adversarial example computed with the fast gradient sign method and the resulting model prediction \ref{f:fs_adv_ex_pred} (see Fig.~\ref{f:model_prediction} for the original mask and model prediction of the same CT image).}\label{f:fast_grad_sign}
\end{figure*}

In practice, for the liver segmentation example this results in a slightly perturbed CT image with a clearly deteriorated model prediction for the liver mask. Fig.~\ref{f:fast_grad_sign} shows the original image from the test set together with the corresponding adversarial example and the resulting model prediction. Compared to the ground truth and the model prediction for the unperturbed CT image in Fig.~\ref{f:model_prediction}, the predicted mask in Fig.~\ref{f:fast_grad_sign} is clearly deteriorated. 

In addition to the fast gradient sign method, many other attack strategies have been developed, such as e.g.\ the targeted attacks from Refs.~\cite{Papernot2016,Carlini2016}, one pixel attacks proposed in Ref.~\cite{Su2019} (and applied to the medical domain in Ref.~\cite{Korpihalkola2020}), the expectation-over-transformation attacks \cite{Athalye2017} or the framework of Ref.~\cite{Zhao2017}. Interestingly, also methods from interpretable AI can be vulnerable to adversarial examples, as shown in Ref.~\cite{Zhang2021}.   

There are several speculations of why adversarial examples exist at all. A first guess might be that the non-linearity of neural networks (NN). However, the authors of Ref.~\cite{Goodfellow2014} argued that most NN are trained in a way that forces them to stay in a mostly linear regime and that this is the reason for being prone to adversarial examples. Others have put forward the explanation that the out-of-sample performance and generalizability are at odds with adversarial robustness \cite{Tsipras2018}. Likely, a better understanding of the dynamics of NN \cite{roberts2022} might provide further insights into the nature of adversarial examples. 

The adversarial example in Fig.~\ref{f:fast_grad_sign} illustrates how ML models are prone to carefully crafted noise, despite showing good out-of-sample performance. 
It is probably fair to say that adversarial examples are ML's equivalent to optical illusions for humans. 
Regarding the practical implementation of an ML system in a clinical workflow, the previous example shows standard performance metrics used in ML do not directly relate to a model's robustness. 

Adversarial examples are carefully crafted noise that perturbs the input. Assuming that in a certain medical application there is no voluntary or maleficent attack of the ML system, one might argue that the susceptibility to adversarial examples does not hamper the clinical implementation. However, even in this case adversarial examples allow us to test and assess the robustness of an ML system under extreme and/or worst case. As pointed out in Ref.~\cite{Carlini2019}, evaluating adversarial robustness of a model can provide some useful insights about the behavior of the ML model even if adversarial attacks can be excluded in a specific case. For risk assessments and QA of ML systems adversarial examples can therefore be an invaluable tool. Furthermore, adversarial examples can hint towards large differences between the human and ML process of decision-making. 

\subsection{Adversarial robustness and defense strategy}\label{s:defense}

Knowing that adversarial examples exist, the next step is, of course, to develop defense strategies that can make a model robust against adversarial attacks. As opposed to the landscape of attack methods, research and development of defenses has received less attention \cite{Carlini2016,Carlini2019} and often new defense strategies are quickly shown to have some blind spots. In particular for safety relevant and medical applications this is quite an unsatisfactory situation. 

Often, attacks are classified according to what is known about the model when designing an adversarial example. The example in Sec.~\ref{s:adv_ex} relies on computing the gradients with respect to the input image. This means that in order to construct this kind of adversarial example full knowledge about the model's architecture, loss function and weights is necessary. This is called a \emph{white box} attack. On the other hand, \emph{black box} attacks refer to the case when there is no or limited knowledge about the model parameters. Of course, there are varying degrees of black box knowledge of a model, e.g. access to predicted probabilities, access to training data, etc. In medical applications a white box situation is unlikely, as the vendors of commercial and prorietary software products usually do not disclose much information to the users/clients. 

An important, but somewhat worrisome property of adversarial examples was found already shortly after their discovery: an adversarial example that is constructed for a specific model, works reasonably well also on other models \cite{Goodfellow2014, Szegedy2013, Liu2016}. I.e. in analogy to the possibilities of transfer learning (see e.g. Refs.~\cite{Pan2010, Torrey2010}), adversarial examples can fool models for which they were not designed. This can make even a completely black box attack quite efficient. 

A simple defense strategy is adversarial training (see Refs.~\cite{Madry2017, Goodfellow2014, Tsipras2018}). Rather than computing the model parameters $\theta$ by minimizing the cost function $\mathcal{C}$ as 

\begin{equation}
  \hat{\theta} \eq \arg \min_{\theta} \mathcal{C}(f_\theta(x), y) \;, 
\end{equation}
we can estimate $\theta$ with 

\begin{equation}\label{eq:robust_train}
  \hat{\theta} \eq \arg \min_\theta \max_{\delta \in \Delta} \mathcal{C}(f_\theta(x+\delta), y) \;, 
\end{equation}
where $\Delta$ is the set of possible perturbations of the input $x$. Roughly speaking, $\Delta$ defines how much noise we can add to the original image in order to produce an adversarial example. The $\min$ $\max$ optimization procedure of Eq.~\eqref{eq:robust_train} can be viewed as a two player game\footnote{The adversarial training in Eq.~\eqref{eq:robust_train} bears similarities to the training of generative adversarial networks (GAN) \cite{Goodfellow2016, Goodfellow2014a}, where the optimization is also set up as a two player game but with one player being a generative model.  
} \cite{Goodfellow2014}. 

The adversarial training of a model in Eq.~\eqref{eq:robust_train} is a form of data augmentation: for every sample $x$ of the training set we can add some perturbation $\delta \in \Delta$ and use it as a new input. It becomes therefore clear that training a robust model requires additional computational resources and the authors of Ref.~\cite{Tsipras2018} pointed out that adversarial training reduces the accuracy of a model. In addition, we might face the difficulty of defining a suitable set $\Delta$ (see e.g.\ Ref.~\cite{Zhang2019} for a discussion of limitations of the method in Eq.~\eqref{eq:robust_train}).

\section{Conceptual considerations for QM of ML}\label{s:concept}

As with any medical device or software where it is not straight forward to quantify its robustness, also with AI we have to find a way to deal with the potential failures of ML systems in clinical practice. In general, stringent national and international regulations assign the responsibility for the correct functioning of a medical device to the vendors. 
As seen e.g. in Ref.~\cite{FDA2021} regulators are taking actions in order to provide a regulatory framework for ML/AI applications in the medical domain\footnote{The website \url{https://aicentral.acrdsi.org/} provides a list of FDA approved AI software}. Of course, the robustness of an ML system needs to be addressed within such a framework. 

We believe that the standard practices of radiation oncology provide useful guiding principle. In view of the potential hazards in the medical domain both the vendors and the users bear some responsibility for assuring a save operation of ML tools in clinical practice. On one hand the devices are certified as medical device, meaning that the vendors' duties are well defined, and on the other hand the users are in charge of a stringent QA program. 

The analogy to radiation oncology, and in particular to proton therapy, can be taken also a step further: a particle accelerator with the corresponding beam line and treatment head is an infinitely complex device with almost uncountable individual components that may fail at any time. Often proton therapy facilities are unique prototypes built in research centers with strong ties to accelerator and high-energy physics research institutes (see e.g.\ \url{https://www.psi.ch/en/protontherapy}) and it is therefore not unusual to employ non-certified equipment or software in a high risk clinical process. In such cases, the clinic takes the full responsibility for the device's risk management and QA. From our perspective, the clinical implementation of ML can profit from the experiences in this field.  

We think that for AI applications the vendors should adhere to best practices, in particular with respect to adversarial robustness, and the users should implement these tools in their QM programs. In particular on the user side, this requires an in-house expertise and should not be outsourced.

\subsection{Best practice for robust ML development}\label{s:best_practice}

For the most part, the literature proposes many more or less ad hoc methods for adversarial attacks and defenses. The authors of Ref.~\cite{Carlini2019} therefore came up with some guiding principles on how to evaluate the robustness for a model and even provide a checklist for model developers.    

A basic requirement for evaluating the robustness of a model is to define a \emph{threat model}. This defines the rules and restrictions that must be respected when constructing an adversarial example. A threat model should encompass the goals (e.g. simple misclassification, targeted attack, etc.), knowledge (e.g. white or black box attack, access to data sets, etc.) and capabilities (size and type of input perturbations, etc.) of an ML system's adversary. Without a threat model, we would have to assume that an adversarial example for our model can be constructed in every possible way. Devising a defense strategy would be impossible and we could not provide any statement about the model's robustness. 

Of course, with a precise definition of the threat model, the robustness of the model is well-defined and in some cases even computable, as discussed in Refs.~\cite{Madry2017, Uesato2018}. In terms of the defense strategy outlined in Sec.~\ref{s:defense} this amounts to defining the allowed set of perturbations $\Delta$. Note that a precise definition of validity of the model has to go along with the threat model. In the medical domain this often means either a restriction to a specific devices (see e.g.\ Ref.~\cite{xue2022a}). 


While the authors of Ref.~\cite{Carlini2019} argue that the source code as well as pre-trained models should be fully accessible, we do not believe that in the medical domain software vendors will fully disclose their products. Nevertheless, we advocate for vendors of a clinical ML system to provide a detailed documentation about the robustness of their system. This includes in particular specifications about the employed threat model as well as robustness tests and measures. It would be highly desirable that the users could have insight in the the risk evaluations of the CE markings or FDA approvals. The shared responsibility between the vendors and users means that this kind information needs to be shared. Otherwise the users cannot implement suitable QM measures.


Some scholars have put forward the necessity of model interpretability for the clinical use of AI (see e.g.\ Refs.~\cite{Reyes2020,Kundu2021,Toja2021,Ghassemi2021}). However, we think that a strict requirement of interpretability might be too restrictive. Also, the degree of interpretability and the deduction of robustness measures therefrom can be quite subjective. Interpretability may, however, play an important role in the risk evaluation and for QM measures as shall be discussed in Sec.~\ref{s:qa_example}.

\subsection{Quality management program for AI users}\label{s:qm4users}

Clinics usually have their workflows mapped in a QM system that is closely related to risk management. Every software tool that is used in a clinical workflow is therefore part of a clinic's risk analysis and QM program (as emphasized also in Ref.~\cite{eclair}). Often devices and software are not explicitly listed or considered in a clinic's risk assessment, as long as they are certified as medical devices. If we want to use an ML system in a clinical workflow, its risks need to be assessed and QA measures need to be defined. 

From the user's perspective, we think that ML tools should be approached in the same way as an external beam radiotherapy facility regarding risk assessment and QA. Radiation oncology has shown how unreliable and very complex systems can fit in a clinical workflow that ensures patient safety at a very high level. This is why we believe that the methodology from the AAPM TG-100 report \cite{AAPMTG100} provides a natural conceptual framework for a clinical workflow that includes AI. 

The technical QA program in radiation oncology clinics is traditionally based on national and international recommendations. Such recommendations focus on assessing all functional performance measures of a device and rarely take into account the whole workflow. Furthermore, these recommendations often have problems to keep up with the rapid technological advances due to the rather lengthy publishing process. 
This is why the TG-100 of the AAPM put forward a risk-based methodology that directs the QA measures in a resource-efficient way, while providing an optimal patient safety \cite{AAPMTG100}. Since it considers a specific clinical workflow as a whole, rather than just functional performance measures, it allows for a swift implementation of new technologies by the user. 

It is important to note that faulty outcomes are often due to issues related to the workflow and not necessarily because of technical failures. Using the wording of Sec.~\ref{s:best_practice}, the threat model for a full clinical workflow should not only cover failures of individual components of the process but the process as a whole. This is one of the strong points of the AAPM TG-100 methodology, as the QM measures are conceived based on the risks of individual steps in the whole workflow. 

The TG-100 methodology relies on three principles for risk assessment and mitigation: process mapping, failure mode and effect analysis (FMEA) and fault trees (FT). These are standard tools in safety-critical industries. 

The process for risk analysis and mitigation outlined in Ref.~\cite{AAPMTG100} involves the following steps. 

\begin{enumerate}
  \item \textbf{Process mapping:} In order to assess the risk of a process, it is useful to start with a graphical representation of the whole process. The level of detail of the process map should be considered carefully. 
  \item \textbf{FMEA:} ``FMEA assesses the likelihood of failures in each step of a process and considers their impact on the final process outcome.''\cite{AAPMTG100} This is sometimes also referred to as a bottom-up approach since the analysis starts from possible failures in the process steps. 
  \item \textbf{FT:} A FT is a useful tool to visualize how failures propagate through the process. One starts with a failure in the process outcome and then identifies all possible hazards that could possibly lead to this failure. 
  \item \textbf{QM program:} Once the risk of the process is assessed with a FMEA and FT, the QM program is set up such that it mitigates the major risks that were identified in the FMEA and FT. 
\end{enumerate}

This procedure should be viewed as an iterative process. Depending on the outcome of the risk assessment, the process map, FMEA, FT and QM program can be adapted and the analysis is repeated until the workflow has an acceptable risk of hazards. Note that this methodology is in line with the recommendations of the Particle Therapy Co-Operative Group (PTCOG) in Ref.~\cite{Flanz2016}, where a combination of top-down and bottom-up risk assessments is viewed as most effectively. 


\section{AAPM TG-100 methodology for ML: a generic example}\label{s:qa_example}

Let us see how we can apply the AAPM TG-100 methodology outlined in Sec.~\ref{s:qm4users} to a clinical workflow that includes an ML system. The example at hand is kept as simple as possible and focuses on the ML related parts of the workflow. 

\subsection{Process mapping}

\begin{figure*}
  \centering 
  \includegraphics[width=0.8\textwidth]{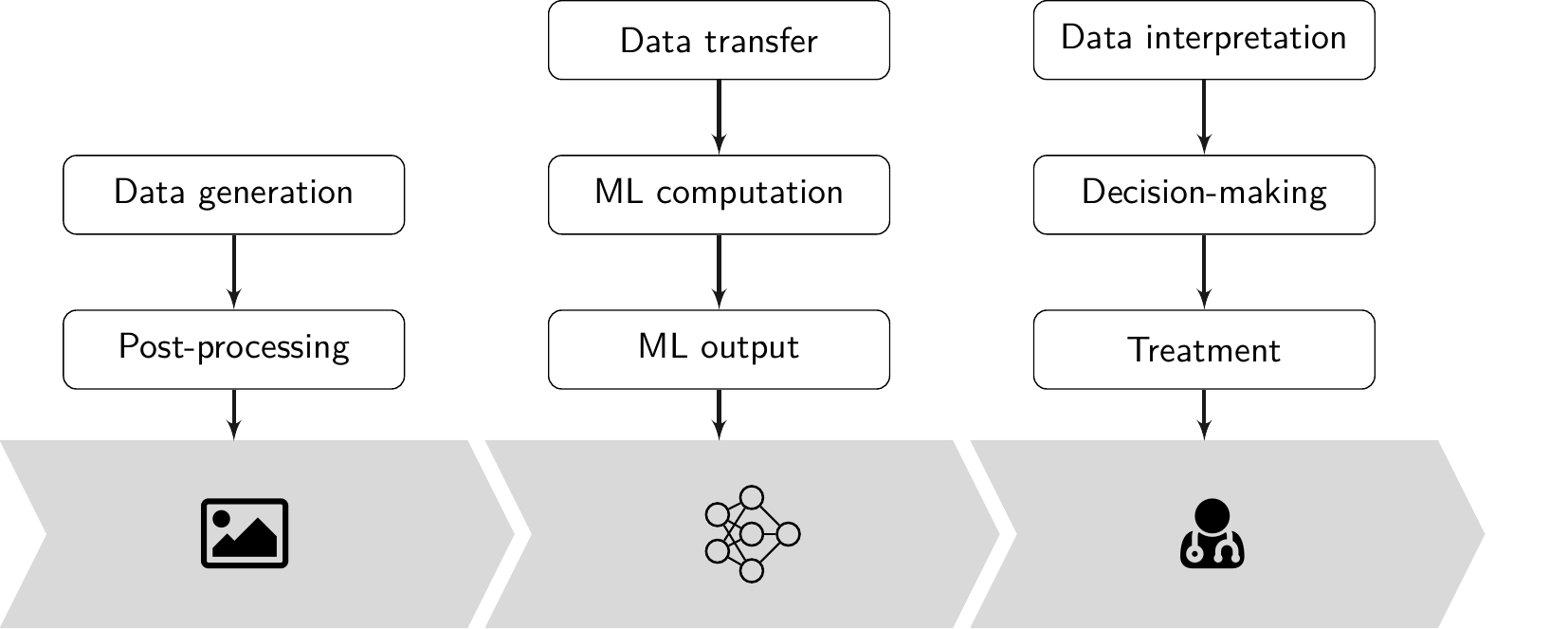}
  \caption{Simplified clinical process with AI support. As a concrete example, one can think of this as a PET/CT image of an oncological patient where the lesion detection is done with ML.}\label{f:process_map}
\end{figure*}

Fig.~\ref{f:process_map} shows a process map, which is the first step towards our risk assessment and QA program design. Our example can be thought of as a generic version of an imaging workflow. First, we generate and post-process the data, as e.g.\ the acquisition of a PET/CT image and the reconstruction of the image. Of course, we omit several steps of a real clinical workflow, such as e.g.\ patient referral or details of the data generation, such as e.g.\ device operation. As discussed in Ref.~\cite{AAPMTG100}, the level of detail of the process map depends strongly on the process itself. 

Next, the data is transferred to an ML system. In real life this step might require quite some attention. We ignore issues related to the actual data transfer, data format, integrity checks, etc. 
The main task in this step is that the ML model computes a certain output. Staying with the example of a PET/CT, the ML system could be a model that automatically segments organs or detects lesions (such as e.g. the models of Refs~\cite{Ermics2020,Zhao2020, Xu2018}).   

In our clinical workflow, we do not allow the ML system to take direct action on the patient or the treatment. The output from the model is interpreted by a physician (or an interdisciplinary board of physicians), who in turn will decide on the further procedures, i.e.\ the ML system is not part of the decision-making process. It should be thought as a support system for the physician. For example, the ML system might be able to speed up significantly a task like lesion detection and segmentation (see e.g. Ref.~\cite{Ermics2020}). Note well that if AI were to take direct and/or automatic action on the patient treatment, the methodology would not change and such a process would still fit in the AAPM TG-100 framework. The risk analysis and QA measures would simply be adapted accordingly.

\subsection{Failure mode and effect analysis}

For every step in the process map, the risk of failure and its consequences are evaluated. As discussed in detail in Ref.~\cite{AAPMTG100}, the FMEA is a prospective risk assessment, i.e.\ the quantification is based on expert knowledge\footnote{Of course, data based assessments are preferable but robust data on failure probabilities, severity, etc. is often not available.}. Applied to AI, this emphasizes the necessity of in-house expertise on AI.  

In a first iteration, the FMEA does not consider any previous QA measures that might already be in place in order not to introduce any bias. Since FMEA is a bottom-up approach, we start by identifying as much failure modes, i.e. ways that each step in our workflow could fail, as possible. Then, the causes and the impact on the final outcome of each failure mode must be determined. Each failure mode is quantified with three figures of merit: occurrence $O$, severity $S$ and lack of detectability $D$. Therefrom, the \emph{risk priority number} $RPN$ is computed as 

\begin{equation}
  RPN \eq O \cdot S \cdot D
\end{equation}

The determination of the values for $O$, $S$, and $D$ is often challenging. Ref.~\cite{AAPMTG100} strongly advises to elaborate the quantification of the FMEA in a cross-professional team. The values of $O$, $S$, and $D$ range from 1 to 10 and correspond to the definition in Tab.\ II of Ref.~\cite{AAPMTG100}. They span the following ranges. 

\begin{itemize}
  \item Occurence $O$: from ``failure unlikely'' (frequency $<0.01\%$) to ``failures inevitable'' (frequency $>5\%$). 
  \item Severity $S$: from ``no effect'' to ``catastrophic''. 
  \item Lack of detectability $D$: from $<10^{-4}$ to $>0.2$ probability of failure being undetected. 
\end{itemize}

In Tab.~\ref{t:fmea} we provide a simplified FMEA for the process shown in Fig.~\ref{f:process_map}. Due to the generic nature of our process, the values of $O$, $S$, and $D$ should not be used as a reference for a real FMEA. Even in a real life example, the often subjective nature of $O$, $S$, and $D$ can induce a large variation.  

\begin{table*}[thb]
  \centering 
  \small 
  \newcolumntype{t}{>{\raggedright\arraybackslash}X}
  %
  %
  %
  %
  \subcaptionbox{ML system.\label{t:fmea_ml}}{
    \begin{tabularx}{0.99\textwidth}{tttcccc}
      \toprule
      Failure mode & Cause & Effects & $O$ & $S$ & $D$ & $RPN$ \\
      \midrule 
      \rowcolor{lightgray} & \textbullet Network failure & & 4 & 4 & 1 & 16 \\
      \rowcolor{lightgray} \multirow[c]{-2}{=}{Faulty data transfer} & \textbullet Wrong data format  &  \multirow[c]{-2}{*}{No or faulty data} & 1 & 6 & 1 & 6  \\
      & \textbullet Hardware failure & & 1 & 5 & 1 & 5 \\
      & \textbullet Non-robust model & & 3 & 8 & 10 & 240 \\
      \multirow[c]{-3}{=}{Faulty model prediction} & \textbullet improper input & \multirow[c]{-3}{*}{Faulty ML output} & 2 & 5 & 3 & 30 \\
      \bottomrule
    \end{tabularx}
  } \\[6ex]
  \subcaptionbox{Interpretation and decision-making.}{
    \begin{tabularx}{0.99\textwidth}{tttcccc}
      \toprule
      Failure mode & Cause & Effects & $O$ & $S$ & $D$ & $RPN$ \\
      \midrule 
      \rowcolor{lightgray} & \textbullet Human error & & 3 & 10 & 4 & 120 \\
      \rowcolor{lightgray} & \textbullet Suboptimal reading conditions & & 2 & 10 & 2 & 40 \\
      \rowcolor{lightgray} \multirow[c]{-3}{=}{Faulty data interpretation} & \textbullet Faulty ML output  &  \multirow[c]{-3}{*}{Patient damage} & 6 & 10 & 2 & 120  \\
       & \textbullet Insufficient decision support & & 3 & 10 & 4 & 120 \\
       \multirow[c]{-2}{=}{Wrong treatment decision} & \textbullet Miscommunication  &  \multirow[c]{-2}{*}{Patient damage} & 2 & 10 & 1 & 20  \\
       \rowcolor{lightgray} & \textbullet Faulty prescription & & 2 & 10 & 2 & 40 \\
       \rowcolor{lightgray} \multirow[c]{-2}{=}{Wrong treatment} & \textbullet Faulty treatment application  &  \multirow[c]{-2}{*}{Patient damage} & 2 & 10 & 1 & 20  \\
      \bottomrule
    \end{tabularx}
  }
  \caption{FMEA risk quantification for the second and third step in the processes of Fig.~\ref{f:process_map}. For brevity's sake, no FMEA for the data generation process is performed.} \label{t:fmea}
\end{table*}

As Tab.~\ref{t:fmea} shows, an ML system can be embedded very smoothly in a FMEA. It is considered simply as a subsystem and/or step in the clinical workflow that takes some input from the previous subprocesses and produces some output that is needed in the subsequent steps. Of course, there are many things that can go wrong in an ML pipeline. We condensed the causes of a faulty model prediction to a ``hardware failure'', an ``unstable model'' and an ``improper input''. 

Nowadays, a hardware failure is fairly rare and the chance to pass undetected is minimal. The consequences might be, however, somewhat severe. Just think about a major delay till the patient can be treated or costs caused by some delay. This is why we assigned the values $O=1$, $S=5$ and $D=1$ which gives a rather low value of $RPN=5$. 

On the other hand, without knowing any details about the ML model's robustness, we have to assume it is inherently vulnerable and unreliable. The occurrence might still be limited with $O=3$, i.e.\ not expecting carefully crafted perturbations of the input, the randomness of data acquisition will not overly strain a model's robustness. However, the severeness can be high as faulty data might lead to wrong treatment decisions and without being detected. Imagine e.g.\ that tumor lesions in a PET/CT image are not detected my the ML system in a situation where the treatment decision is then taken without a review of the images by an experienced nuclear medicine specialist. Therefore, we need to assign a high severity of $S=8$ and a low probability to detect the error $D=10$. Note that so far we could simply make a conservative choice and assume that the ML system lacks robustness. At this point there is no need to rely on any robustness measures from the vendor or e.g. interpretability (as demanded e.g.\ in Ref.~\cite{Kundu2021}). Of course, these figures might be lowered depending on mitigation measures, as we shall discuss in the following. 

Finally, a faulty model output might be produced because the input data is outside the model's validity. Assuming that the data pipeline is more or less robust, e.g.\ the model will not receive an MR image when it expects PET/CT data, the occurrence should be low $O=2$. The severity might be higher $S=5$, but we expect the model to produce an output that is more or less easily detected as faulty, hence $D=3$.  

With the $RPN$ values of Tab.~\ref{t:fmea} we can now identify the process steps that might need some control or revision. 

It is important to be aware of the limitations of the FMEA (see e.g. Ref.~\cite{Flanz2016}). On one hand, if the clinical workflow under consideration is complex 
a full FMEA can be time-consuming. This is particularly true in situations where the quantification is done by a cross-professional committee. However, this is not the only issue. The quantification of a risk in terms of $O$, $S$ and $D$ is subjective, since reliable empirical data is often missing. This makes the $RPN$ a rather uncertain figure of merit with possibly large variations. Furthermore, the $RPN$ might not reflect the true risk (is it meaningful to simply multiply $O$, $S$ and $D$?) and there are issues related to the distribution of the numerical values (see Ref.~\cite{Buchgeister2021}).

\subsection{Fault tree}

\begin{figure*}[tbh]
  \centering 
  \includegraphics[width=0.85\textwidth]{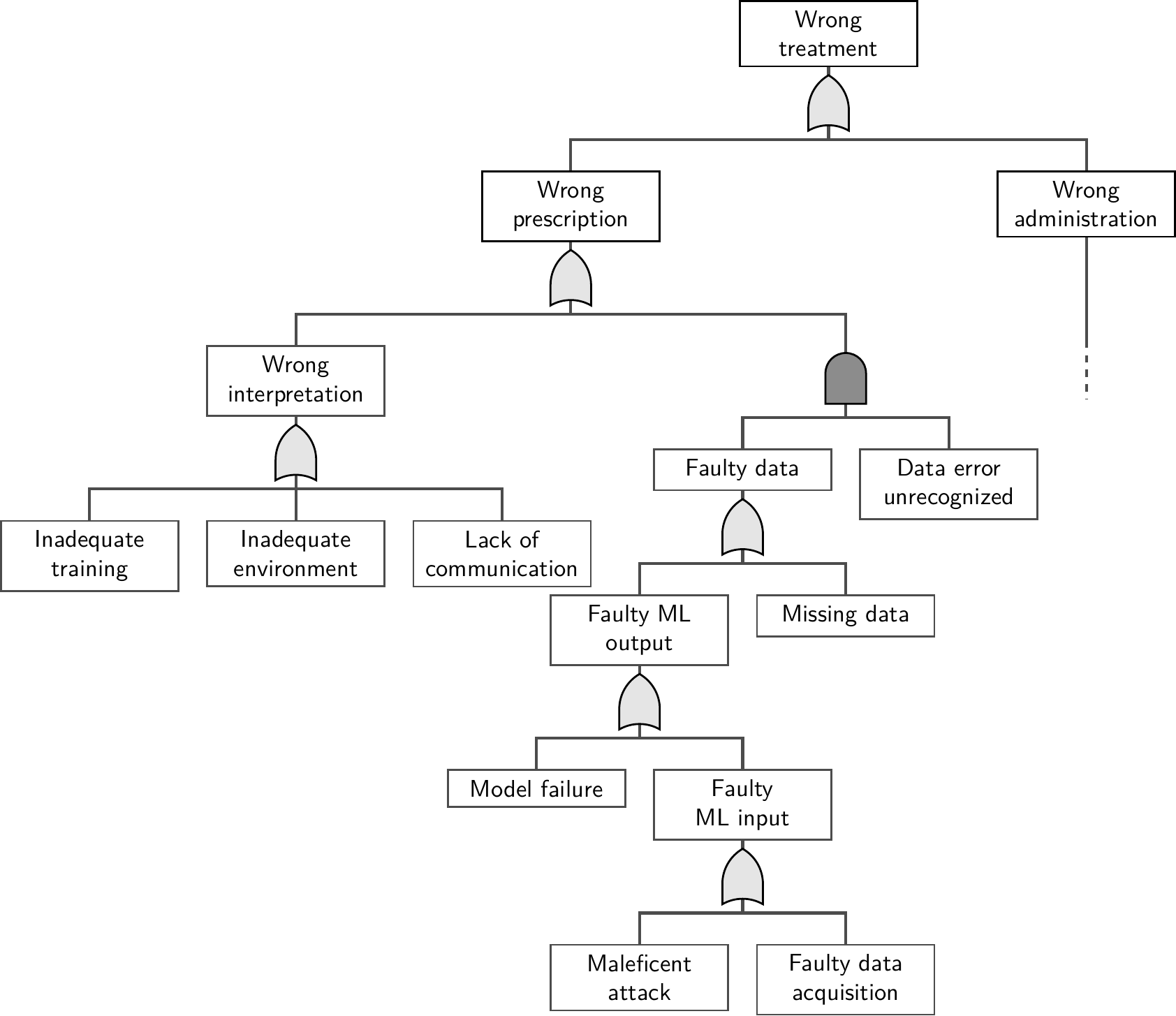}
  \caption{Simplified FT for the process of Fig.~\ref{f:process_map}. The symbol \includegraphics[width=8pt]{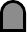} represents an ``and'' gate while \includegraphics[width=8pt]{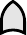} is an ``or'' gate. }\label{f:ft}
\end{figure*}

The FT analysis complements the FMEA and helps to uncover risks, and in particular interconnection between process steps, that might be somewhat hidden in the FMEA. In Fig.~\ref{f:ft} we show a simplified FT for the process in Fig.~\ref{f:process_map}, omitting the continuation of certain branches in the FT in order to keep it simple and focussed on AI. 

The FT analysis starts with the error at the end of the process, which in our example is the wrong treatment of the patient. Then one has to find all possible sources in the workflow that might lead to this hazard. It then depends on whether there are multiple factors that need to be satisfied in order to produce a failure (logical ``and'' gate) or if a single factor can lead itself to an error (logical ``or'' gate). As seen in Fig.~\ref{f:ft}, the wrong treatment can be caused either by a wrong prescription or by a wrong administration of the treatment. It is then clear that ``and'' gates represent a safety feature, since multiple conditions must be met in order to produce a failure. On the other hand, ``or'' gates should be investigated more closely since they bear the risk of error propagation. Therefore, any QA measure should start at an ``or'' gates. 

The ML related branch of the FT is fairly simple. There are two possibilities that can produce a wrong or incomplete model output: either there is a problem with the input or with the model itself. The input could be faulty or incomplete e.g.\ because of random variations in the data acquisition, wrong data format, wrong data protocol, incomplete data transfer, adversarial attacks etc.\footnote{In order to be thorough, we should continue the FT and cover also the data acquisition and address at least some of these possible problems.} Regarding a failure of the model, imagine e.g.\ that the model is not adequate for the task or that the model performance is not as expected. 

The most important part of the FT in Fig.~\ref{f:ft} is the ``and'' gate that connects the faulty output from the ML model with the event that the error is not recognized by the physicians in the decision-making process. Remember that in our example from Fig.~\ref{f:process_map}, the output of the ML system is fed to the decision-making process. Physicians will review the data and use it basis for their decision. Hence, a wrong prescription will only occur if the ML output is incorrect and the error remains unrecognized in the decision-making process. Due to this ``and'' gate, we could in principle focus all our QA measures on the decision-making process and ensure that every ML output error gets recognized. This allows us to compensate for the high $RPN$ of the ML system in the FMEA (see Tab.~\ref{t:fmea_ml}). Or in other words, in a second iteration of the FMEA we would reduce the value of $D$ due to the ``and'' gate in our FT. 

Note that also a FT has some limitations. E.g.\ it can become difficult to visualize and account for complex interactions between the levels of the FT (see Ref.~\cite{Flanz2016}).

\subsection{QA program}

Based on the process map, the FMEA and the FT we could now design the risk mitigation measures and the QA program for our clinical workflow. The FMEA and the risk quantification show which steps are most in need of risk mitigation measures. On the other hand, the FT give some insights in where the propagation of errors is most efficiently blocked. If a single QM step or measure is sufficient to block an error from creating an incident depends on the specific case. In general, however, it is advisable to have multiple ``barriers'' preventing the propagation of errors, as discussed in Ref.~\cite{AAPMTG100}. 
Having multiple QM measures that can block an error will certainly reduce the value of $RPN$, mostly because of a reduction of $D$. 

Considering our FMEA, it is clear that we should focus our QM efforts on the ML system's lack of robustness, the data interpretation, the decision-making process and the problems related to contraindication (see Tab.~\ref{t:fmea}). From the FT, we know that faults in the ML output could be compensated by a robust data interpretation. In a second iteration of the analysis, we could therefore e.g.\ lower the value of $D$ in the FMEA if our QA measures in the decision-making and prescription process can prevent a faulty ML output to propagate further in the FT. This example illustrates nicely how even an unreliable ML system can be implemented in a robust and safe clinical workflow. Nevertheless, it would be only a single step that prevents a ML failure from producing a hazard for the patient. In the next section we therefore outline some possible approaches for QM of an ML system that can help to reduce the $RPN$'s dependence on subsequent steps.

\subsection{Risk mitigation for ML systems}

What possibilities do we have for mitigating the risks of an ML model? As mentioned before, we believe it is the user's responsibility to address the possible hazards and the AAPM TG-100 \cite{AAPMTG100} methodology provides the ideal framework. In Sec.~\ref{s:qa_example} we saw how the data interpretation and decision-making process can make up for the ML system's lack of robustness. Depending on the involved risk, it might however not be enough to solely rely on a single QM measure.  

While vendors should adhere as much as possible to the principles discussed in Sec.~\ref{s:best_practice} (see Ref.~\cite{Carlini2019}), a clinic must have the knowledge and resources to understand and address possible hazards of an ML system. The best practices discussed in Sec.~\ref{s:best_practice}, of course have, an influence on the risk scores in the FMEA. Consider e.g.\ that a vendor's threat model has been tailored to a specific task in a clinical workflow and that it covers a wide range of failures in that workflow. This would allow the user to reduce $O$ and possibly $D$ of the ML system in the FMEA. 

In analogy to the recurring dosimetric measurements of a QA program in radiation oncology, we believe that setting up a series of periodic in-house and vendor independent tests of the ML system's performance can provide confidence and trust in the ML model. This could be e.g. a clinic specific test data set that includes specifically designed adversarial examples, randomly generated data, corrupted or otherwise faulty input, etc. It is important to keep in mind that such an in-house test should cover as much as possible the intended use of the software, as defined e.g.\ in the CE labeling or FDA approval of the software. This would also allow for an analysis of the whole data pipeline and assessment of the effectiveness of data consistency checks through the whole clinical workflow. Furthermore, such an in-house test data set allows to assess changes in the model whenever there is a software update or if the ML system is itself adaptive and learns from its routine operation.   

Another important aspect for increasing the robustness of an AI model is its interpretability (see e.g.\ Ref.\cite{molnar2022} for a general introduction to the topic). The basic idea of interpretable ML is to find models and/or develop methods that allow for a human interpretation or explanation of the model's output. Some scholars have recently argued that interpretability should be a requirement for ML systems in medical applications (see Refs.~\cite{Kundu2021,Toja2021,Ghassemi2021,Liu2021}). We are convinced that interpretability can play a major role in assessing the possible risks of a ML model and thereby lower the values of $O$ and $S$ in the FMEA. 
Note however, that Ref.~\cite{Zhang2021} constructed adversarial examples for the interpretation of models, i.e.\ even what is perceived as an interpretation of a model seems to have some inherent vulnerability and/or lack of robustness. Furthermore, the authors of Ref.~\cite{Ghassemi2021} show how current interpretation/explanation methods might fail at providing decision support (see also Ref.~\cite{Arun2021}). 

Our perspective allows us to be somewhat agnostic towards interpretability. Implementing an interpretable ML model can alter the risk assessment, i.e. the $RPN$, and establish confidence in the model's performance. The same applies to in-house test data sets and best practices when constructing the model (see Sec.~\ref{s:best_practice}). The key advantage of the AAPM TG-100 \cite{AAPMTG100} methodology is that it does not matter how well a model is interpretable or how robust it is. It is always possible to implement it in a clinical workflow. The whole workflow and the QM measures will then simply be adapted according to the risk that the ML model represents.

\section{Conclusions}

Despite the huge potential benefits that AI can bring to healthcare, there are some fundamental issues. One of the mayor concerns for the clinical implementation of AI is the robustness of such tools. In this paper, we propose a conceptual framework and discuss how an AI system can be implemented in a clinical workflow. 

We exemplify with a simple segmentation example how easy it is to construct adversarial examples that force an ML model to perform poorly despite having good out-of-sample figures of merit. Such carefully crafted noise, which is added to the model input, is hardly perceivable by the human eye. The existence of adversarial examples for ML systems raises concerns for the use of such systems in high risk environments like healthcare. However, adversarial examples can provide unique stress tests can be very useful to assess a model's robustness and worst case stability.

The clinical implementation of AI tools is in strong need of a conceptual framework where the robustness issues of AI can be addressed systematically. We argue that AI in healthcare should draw from the field of radiation oncology, where complex and possibly unreliable equipment that can cause high patient damage is being used in clinical routine. First, we take the view that there is a shared responsibility between the vendors and users when implementing ML system in a clinical workflow. On one hand, the vendors (or developers) should adhere to best practices and, most importantly, should disclose how they address the robustness of their systems. On the other hand, the users are responsible for implementing the ML system in a QM system and setting up appropriate QA measures. For both, the assessment of a model's robustness is crucial and adversarial examples play a key role in tackling these questions.

We advocate to use the methodology from the AAPM Task Group 100 report \cite{AAPMTG100} develop a QA program for complex clinical processes that include ML systems. With a generic example of an imaging workflow we illustrate this point by performing a process map, FMEA and FTA. One of the key points of the AAPM Task Group 100 \cite{AAPMTG100} framework is that the whole clinical process is considered in the risk assessment and therefore allows for an efficient construction of a QA program. Also, this methodology does not depend on reliability of an ML system. Rather, it provides a framework that can accommodate any level of AI robustness. Of course, the risk evaluation will, change according to a model's robustness and therefore also the necessary QA measures can be adapted. We stress that interpretability of a ML model is not a strict requirement in this framework, but becomes a risk mitigation strategy that can significantly reduce the risk scores in the FMEA.

\bibliography{qa_for_ml_project_bibliography}

\begin{thebibliography}{10}
\expandafter\ifx\csname url\endcsname\relax
  \def\url#1{\texttt{#1}}\fi
\expandafter\ifx\csname urlprefix\endcsname\relax\def\urlprefix{URL }\fi
\expandafter\ifx\csname href\endcsname\relax
  \def\href#1#2{#2} \def\path#1{#1}\fi

\bibitem{AAPMTG100}
M.~S. Huq, B.~A. Fraass, P.~B. Dunscombe, J.~P. Gibbons~Jr., G.~S. Ibbott,
  A.~J. Mundt, S.~Mutic, J.~R. Palta, F.~Rath, B.~R. Thomadsen, J.~F.
  Williamson, E.~D. Yorke, The report of {Task Group 100 of the AAPM}:
  Application of risk analysis methods to radiation therapy quality management,
  Medical Physics 43~(7) (2016) 4209--4262.
\newblock \href {https://doi.org/10.1118/1.4947547}
  {\path{doi:10.1118/1.4947547}}.

\bibitem{Giger2018}
M.~L. Giger, Machine learning in medical imaging, Journal of the American
  College of Radiology 15~(3, Part B) (2018) 512 -- 520, data Science: Big Data
  Machine Learning and Artificial Intelligence.
\newblock \href {https://doi.org/https://doi.org/10.1016/j.jacr.2017.12.028}
  {\path{doi:https://doi.org/10.1016/j.jacr.2017.12.028}}.

\bibitem{Sahiner2019}
B.~Sahiner, A.~Pezeshk, L.~M. Hadjiiski, X.~Wang, K.~Drukker, K.~H. Cha, R.~M.
  Summers, M.~L. Giger, Deep learning in medical imaging and radiation therapy,
  Medical Physics 46~(1) (2019) e1--e36.
\newblock \href {https://doi.org/https://doi.org/10.1002/mp.13264}
  {\path{doi:https://doi.org/10.1002/mp.13264}}.

\bibitem{Shen2020}
C.~Shen, D.~Nguyen, Z.~Zhou, S.~B. Jiang, B.~Dong, X.~Jia, An introduction to
  deep learning in medical physics: advantages, potential, and challenges,
  Physics in Medicine {\&} Biology 65~(5) (2020) 05TR01.
\newblock \href {https://doi.org/10.1088/1361-6560/ab6f51}
  {\path{doi:10.1088/1361-6560/ab6f51}}.

\bibitem{Maier2019}
A.~Maier, C.~Syben, T.~Lasser, C.~Riess, A gentle introduction to deep learning
  in medical image processing, Zeitschrift f{\"u}r Medizinische Physik 29~(2)
  (2019) 86--101.

\bibitem{Madabhushi2016}
A.~Madabhushi, G.~Lee, Image analysis and machine learning in digital
  pathology: Challenges and opportunities, Medical Image Analysis 33 (2016) 170
  -- 175, 20th anniversary of the Medical Image Analysis journal (MedIA).
\newblock \href {https://doi.org/10.1016/j.media.2016.06.037}
  {\path{doi:10.1016/j.media.2016.06.037}}.

\bibitem{Lustberg2018}
T.~Lustberg, J.~{van Soest}, M.~Gooding, D.~Peressutti, P.~Aljabar, J.~{van der
  Stoep}, W.~{van Elmpt}, A.~Dekker, Clinical evaluation of atlas and deep
  learning based automatic contouring for lung cancer, Radiotherapy and
  Oncology 126~(2) (2018) 312--317.
\newblock \href {https://doi.org/10.1016/j.radonc.2017.11.012}
  {\path{doi:10.1016/j.radonc.2017.11.012}}.

\bibitem{Ermics2020}
E.~Ermi{\c{s}}, A.~Jungo, R.~Poel, M.~Blatti-Moreno, R.~Meier, U.~Knecht, D.~M.
  Aebersold, M.~K. Fix, P.~Manser, M.~Reyes, E.~Herrmann, Fully automated brain
  resection cavity delineation for radiation target volume definition in
  glioblastoma patients using deep learning, Radiation oncology 15~(100) (2020)
  1--10.

\bibitem{Wang2015}
T.~Wang, L.~Cao, W.~Yang, Q.~Feng, W.~Chen, Y.~Zhang, {Adaptive patch-based
  {POCS} approach for super resolution reconstruction of 4D-{CT} lung data},
  Physics in Medicine and Biology 60~(15) (2015) 5939--5954.
\newblock \href {https://doi.org/10.1088/0031-9155/60/15/5939}
  {\path{doi:10.1088/0031-9155/60/15/5939}}.

\bibitem{Xue2021}
S.~Xue, R.~Guo, K.~P. Bohn, J.~Matzke, M.~Viscione, I.~Alberts, H.~Meng,
  C.~Sun, M.~Zhang, M.~Zhang, R.~Sznitman, G.~El~Fakhri, A.~Rominger, B.~Li,
  K.~Shi, A cross-scanner and cross-tracer deep learning method for the
  recovery of standard-dose imaging quality from low-dose pet., European
  journal of nuclear medicine and molecular imaging (Dec. 2021).
\newblock \href {https://doi.org/10.1007/s00259-021-05644-1}
  {\path{doi:10.1007/s00259-021-05644-1}}.

\bibitem{Hong2018}
X.~Hong, Y.~Zan, F.~Weng, W.~Tao, Q.~Peng, Q.~Huang, {Enhancing the Image
  Quality via Transferred Deep Residual Learning of Coarse PET Sinograms}, IEEE
  Transactions on Medical Imaging 37~(10) (2018) 2322--2332.
\newblock \href {https://doi.org/10.1109/TMI.2018.2830381}
  {\path{doi:10.1109/TMI.2018.2830381}}.

\bibitem{Xu2022}
J.~Xu, F.~Noo, {Convex optimization algorithms in medical image
  reconstruction-in the age of AI.}, Physics in medicine and biology 67 (Mar.
  2022).
\newblock \href {https://doi.org/10.1088/1361-6560/ac3842}
  {\path{doi:10.1088/1361-6560/ac3842}}.

\bibitem{Wang2018}
G.~Wang, J.~C. Ye, K.~Mueller, J.~A. Fessler, {Image Reconstruction is a New
  Frontier of Machine Learning}, IEEE Transactions on Medical Imaging 37~(6)
  (2018) 1289--1296.
\newblock \href {https://doi.org/10.1109/TMI.2018.2833635}
  {\path{doi:10.1109/TMI.2018.2833635}}.

\bibitem{Zhao2020}
Y.~Zhao, A.~Gafita, B.~Vollnberg, G.~Tetteh, F.~Haupt, A.~Afshar-Oromieh,
  B.~Menze, M.~Eiber, A.~Rominger, K.~Shi, {Deep neural network for automatic
  characterization of lesions on Ga-PSMA-11 PET/CT}, European journal of
  nuclear medicine and molecular imaging 47 (2020) 603--613.
\newblock \href {https://doi.org/10.1007/s00259-019-04606-y}
  {\path{doi:10.1007/s00259-019-04606-y}}.

\bibitem{Xu2018}
L.~Xu, G.~Tetteh, J.~Lipkova, Y.~Zhao, H.~Li, P.~Christ, M.~Piraud, A.~Buck,
  K.~Shi, B.~H. Menze, {Automated Whole-Body Bone Lesion Detection for Multiple
  Myeloma on , javax.xml.bind.JAXBElement@1f89384c, Ga-Pentixafor PET/CT
  Imaging Using Deep Learning Methods.}, Contrast media \& molecular imaging
  2018 (2018) 2391925.
\newblock \href {https://doi.org/10.1155/2018/2391925}
  {\path{doi:10.1155/2018/2391925}}.

\bibitem{Xue2020}
S.~Xue, A.~Gafita, A.~Afshar-Oromieh, M.~Eiber, A.~Rominger, K.~Shi, Voxel-wise
  prediction of post-therapy dosimetry for 177lu-psma i\&t therapy using deep
  learning, Journal of Nuclear Medicine 61~(supplement 1) (2020) 1424--1424.

\bibitem{Wang2020}
M.~Wang, Q.~Zhang, S.~Lam, J.~Cai, R.~Yang, A review on application of deep
  learning algorithms in external beam radiotherapy automated treatment
  planning., Frontiers in oncology 10 (2020) 580919.
\newblock \href {https://doi.org/10.3389/fonc.2020.580919}
  {\path{doi:10.3389/fonc.2020.580919}}.

\bibitem{Xue2022}
S.~Xue, A.~Gafita, C.~Dong, Y.~Zhao, G.~Tetteh, B.~H. Menze, S.~Ziegler,
  W.~Weber, A.~Afshar-Oromieh, A.~Rominger, M.~Eiber, K.~Shi, {Proof-of-concept
  Study to Estimate Individual Post-Therapy Dosimetry in Men with Advanced
  Prostate Cancer Treated with 177Lu-PSMA I{\&}T Therapy} (2022).
\newblock \href {https://doi.org/10.21203/rs.3.rs-1588151/}
  {\path{doi:10.21203/rs.3.rs-1588151/}}.

\bibitem{FDA2021}
{US Food and Drug Administration}, et~al., Artificial intelligence/machine
  learning (ai/ml)--based software as a medical device (samd) action plan (Jan.
  2021).

\bibitem{Muehlematter2021}
U.~J. Muehlematter, P.~Daniore, K.~N. Vokinger, {Approval of artificial
  intelligence and machine learning-based medical devices in the USA and Europe
  (2015–20): a comparative analysis}, The Lancet Digital Health 3~(3) (2021)
  e195--e203.
\newblock \href {https://doi.org/10.1016/S2589-7500(20)30292-2}
  {\path{doi:10.1016/S2589-7500(20)30292-2}}.

\bibitem{Hwang2019}
T.~J. Hwang, A.~S. Kesselheim, K.~N. Vokinger, {Lifecycle Regulation of
  Artificial Intelligence- And Machine Learning-Based Software Devices in
  Medicine}, JAMA - Journal of the American Medical Association 322~(23) (2019)
  2285 – 2286, cited by: 36.
\newblock \href {https://doi.org/10.1001/jama.2019.16842}
  {\path{doi:10.1001/jama.2019.16842}}.

\bibitem{Benjamens2020}
S.~Benjamens, P.~Dhunnoo, B.~Meskó, {The state of artificial
  intelligence-based FDA-approved medical devices and algorithms: an online
  database}, npj Digital Medicine 3~(1), cited by: 153; All Open Access, Gold
  Open Access, Green Open Access (2020).
\newblock \href {https://doi.org/10.1038/s41746-020-00324-0}
  {\path{doi:10.1038/s41746-020-00324-0}}.

\bibitem{Paschali2018}
M.~Paschali, S.~Conjeti, F.~Navarro, N.~Navab, {Generalizability vs.
  Robustness: Adversarial Examples for Medical Imaging} (Mar. 2018).
\newblock \href {http://arxiv.org/abs/1804.00504} {\path{arXiv:1804.00504}}.

\bibitem{Flanz2016}
J.~Flanz, O.~J{\"a}kel, E.~Ford, S.~Hahn, A.~Mazal, J.~Daartz, {PTCOG Safety
  Group Report on Aspects of Safety in Particle Therapy, Version 2} (May 2016).

\bibitem{Szegedy2013}
C.~Szegedy, W.~Zaremba, I.~Sutskever, J.~Bruna, D.~Erhan, I.~Goodfellow,
  R.~Fergus, Intriguing properties of neural networks (Dec. 2013).
\newblock \href {http://arxiv.org/abs/1312.6199v4} {\path{arXiv:1312.6199v4}}.

\bibitem{Akhtar2018}
N.~Akhtar, A.~Mian, Threat of adversarial attacks on deep learning in computer
  vision: A survey, IEEE Access 6 (2018) 14410--14430.
\newblock \href {https://doi.org/10.1109/ACCESS.2018.2807385}
  {\path{doi:10.1109/ACCESS.2018.2807385}}.

\bibitem{Biggio2018}
B.~Biggio, F.~Roli, Wild patterns: Ten years after the rise of adversarial
  machine learning, Pattern Recognition 84 (2018) 317--331.
\newblock \href {https://doi.org/10.1016/j.patcog.2018.07.023}
  {\path{doi:10.1016/j.patcog.2018.07.023}}.

\bibitem{Miller2020}
D.~J. Miller, Z.~Xiang, G.~Kesidis, Adversarial learning targeting deep neural
  network classification: A comprehensive review of defenses against attacks,
  Proceedings of the IEEE 108~(3) (2020) 402--433.
\newblock \href {https://doi.org/10.1109/JPROC.2020.2970615}
  {\path{doi:10.1109/JPROC.2020.2970615}}.

\bibitem{Jin2020}
W.~Jin, Y.~Li, H.~Xu, Y.~Wang, S.~Ji, C.~Aggarwal, J.~Tang, Adversarial attacks
  and defenses on graphs: A review, a tool and empirical studies (Mar. 2020).
\newblock \href {http://arxiv.org/abs/2003.00653} {\path{arXiv:2003.00653}}.

\bibitem{Xu2020}
H.~Xu, Y.~Ma, H.-C. Liu, D.~Deb, H.~Liu, J.-L. Tang, A.~K. Jain, Adversarial
  attacks and defenses in images, graphs and text: A review, International
  Journal of Automation and Computing 17~(2) (2020) 151--178.

\bibitem{Liu2021}
N.~Liu, M.~Du, R.~Guo, H.~Liu, X.~Hu, Adversarial attacks and defenses: An
  interpretation perspective (2021) 86–99\href
  {https://doi.org/10.1145/3468507.3468519}
  {\path{doi:10.1145/3468507.3468519}}.

\bibitem{Ronneberger2015}
O.~Ronneberger, P.~Fischer, T.~Brox, U-net: Convolutional networks for
  biomedical image segmentation, in: N.~Navab, J.~Hornegger, W.~M. Wells, A.~F.
  Frangi (Eds.), Medical Image Computing and Computer-Assisted Intervention --
  MICCAI 2015, Springer International Publishing, Cham, 2015, pp. 234--241.
\newblock \href {http://arxiv.org/abs/1505.04597} {\path{arXiv:1505.04597}}.

\bibitem{Bilic2019}
P.~Bilic, P.~F. Christ, E.~Vorontsov, G.~Chlebus, H.~Chen, Q.~Dou, C.-W. Fu,
  X.~Han, P.-A. Heng, J.~Hesser, S.~Kadoury, T.~Konopczynski, M.~Le, C.~Li,
  X.~Li, J.~Lipkovà, J.~Lowengrub, H.~Meine, J.~H. Moltz, C.~Pal, M.~Piraud,
  X.~Qi, J.~Qi, M.~Rempfler, K.~Roth, A.~Schenk, A.~Sekuboyina, E.~Vorontsov,
  P.~Zhou, C.~Hülsemeyer, M.~Beetz, F.~Ettlinger, F.~Gruen, G.~Kaissis,
  F.~Lohöfer, R.~Braren, J.~Holch, F.~Hofmann, W.~Sommer, V.~Heinemann,
  C.~Jacobs, G.~E.~H. Mamani, B.~van Ginneken, G.~Chartrand, A.~Tang,
  M.~Drozdzal, A.~Ben-Cohen, E.~Klang, M.~M. Amitai, E.~Konen, H.~Greenspan,
  J.~Moreau, A.~Hostettler, L.~Soler, R.~Vivanti, A.~Szeskin, N.~Lev-Cohain,
  J.~Sosna, L.~Joskowicz, B.~H. Menze, The liver tumor segmentation benchmark
  (lits) (Jan. 2019).
\newblock \href {http://arxiv.org/abs/1901.04056} {\path{arXiv:1901.04056}}.

\bibitem{Lin2017}
T.-Y. Lin, P.~Goyal, R.~Girshick, K.~He, P.~Dollár, Focal loss for dense
  object detection (Aug. 2017).
\newblock \href {http://arxiv.org/abs/1708.02002} {\path{arXiv:1708.02002}}.

\bibitem{Sorensen1948}
T.~A. S{\o}rensen, A method of establishing groups of equal amplitude in plant
  sociology based on similarity of species content and its application to
  analyses of the vegetation on danish commons, Biol. Skar. 5 (1948) 1--34.

\bibitem{Dice1945}
L.~R. Dice, Measures of the amount of ecologic association between species,
  Ecology 26~(3) (1945) 297--302.
\newblock \href {https://doi.org/https://doi.org/10.2307/1932409}
  {\path{doi:https://doi.org/10.2307/1932409}}.

\bibitem{Goodfellow2014}
I.~J. Goodfellow, J.~Shlens, C.~Szegedy, Explaining and harnessing adversarial
  examples (Dec. 2014).
\newblock \href {http://arxiv.org/abs/1412.6572v3} {\path{arXiv:1412.6572v3}}.

\bibitem{Papernot2016}
N.~Papernot, P.~McDaniel, S.~Jha, M.~Fredrikson, Z.~B. Celik, A.~Swami, The
  limitations of deep learning in adversarial settings, in: 2016 IEEE European
  Symposium on Security and Privacy (EuroS P), 2016, pp. 372--387.
\newblock \href {https://doi.org/10.1109/EuroSP.2016.36}
  {\path{doi:10.1109/EuroSP.2016.36}}.

\bibitem{Carlini2016}
N.~Carlini, D.~Wagner, Towards evaluating the robustness of neural networks
  (Aug. 2016).
\newblock \href {http://arxiv.org/abs/1608.04644} {\path{arXiv:1608.04644}}.

\bibitem{Su2019}
J.~Su, D.~V. Vargas, K.~Sakurai, One pixel attack for fooling deep neural
  networks, IEEE Transactions on Evolutionary Computation 23 (2019) 828--841.
\newblock \href {https://doi.org/10.1109/TEVC.2019.2890858}
  {\path{doi:10.1109/TEVC.2019.2890858}}.

\bibitem{Korpihalkola2020}
J.~Korpihalkola, T.~Sipola, S.~Puuska, T.~Kokkonen, One-pixel attack deceives
  computer-assisted diagnosis of cancer, 2021 4th International Conference on
  Signal Processing and Machine Learning (SPML 2021) (2021) 100-106 (Dec.
  2020).
\newblock \href {http://arxiv.org/abs/2012.00517} {\path{arXiv:2012.00517}},
  \href {https://doi.org/10.1145/3483207.3483224}
  {\path{doi:10.1145/3483207.3483224}}.

\bibitem{Athalye2017}
A.~Athalye, L.~Engstrom, A.~Ilyas, K.~Kwok, Synthesizing robust adversarial
  examples (Jul. 2017).
\newblock \href {http://arxiv.org/abs/1707.07397} {\path{arXiv:1707.07397}}.

\bibitem{Zhao2017}
Z.~Zhao, D.~Dua, S.~Singh, Generating natural adversarial examples (Oct. 2017).
\newblock \href {http://arxiv.org/abs/1710.11342} {\path{arXiv:1710.11342}}.

\bibitem{Zhang2021}
J.~Zhang, H.~Chao, M.~K. Kalra, G.~Wang, P.~Yan, Overlooked trustworthiness of
  explainability in medical ai, medRxiv (2021).
\newblock \href {https://doi.org/10.1101/2021.12.23.21268289}
  {\path{doi:10.1101/2021.12.23.21268289}}.

\bibitem{Tsipras2018}
D.~Tsipras, S.~Santurkar, L.~Engstrom, A.~Turner, A.~Madry, Robustness may be
  at odds with accuracy (May 2018).
\newblock \href {http://arxiv.org/abs/1805.12152} {\path{arXiv:1805.12152}}.

\bibitem{roberts2022}
D.~A. Roberts, S.~Yaida, B.~Hanin, Frontmatter, Cambridge University Press,
  2022.
\newblock \href {http://arxiv.org/abs/2106.10165} {\path{arXiv:2106.10165}}.

\bibitem{Carlini2019}
N.~Carlini, A.~Athalye, N.~Papernot, W.~Brendel, J.~Rauber, D.~Tsipras,
  I.~Goodfellow, A.~Madry, A.~Kurakin, On evaluating adversarial robustness
  (Feb. 2019).
\newblock \href {http://arxiv.org/abs/1902.06705v2}
  {\path{arXiv:1902.06705v2}}.

\bibitem{Liu2016}
Y.~Liu, X.~Chen, C.~Liu, D.~Song, Delving into transferable adversarial
  examples and black-box attacks (Nov. 2016).
\newblock \href {http://arxiv.org/abs/1611.02770} {\path{arXiv:1611.02770}}.

\bibitem{Pan2010}
S.~J. Pan, Q.~Yang, A survey on transfer learning, IEEE Transactions on
  Knowledge and Data Engineering 22~(10) (2010) 1345--1359.
\newblock \href {https://doi.org/10.1109/TKDE.2009.191}
  {\path{doi:10.1109/TKDE.2009.191}}.

\bibitem{Torrey2010}
L.~Torrey, J.~Shavlik, Transfer learning, in: Handbook of research on machine
  learning applications and trends: algorithms, methods, and techniques, IGI
  global, 2010, pp. 242--264.

\bibitem{Madry2017}
A.~Madry, A.~Makelov, L.~Schmidt, D.~Tsipras, A.~Vladu, Towards deep learning
  models resistant to adversarial attacks (Jun. 2017).
\newblock \href {http://arxiv.org/abs/1706.06083} {\path{arXiv:1706.06083}}.

\bibitem{Goodfellow2016}
I.~Goodfellow, Nips 2016 tutorial: Generative adversarial networks (Dec. 2016).
\newblock \href {http://arxiv.org/abs/1701.00160v4}
  {\path{arXiv:1701.00160v4}}.

\bibitem{Goodfellow2014a}
I.~J. Goodfellow, J.~Pouget{-}Abadie, M.~Mirza, B.~Xu, D.~Warde{-}Farley,
  S.~Ozair, A.~C. Courville, Y.~Bengio, Generative adversarial nets, in:
  Z.~Ghahramani, M.~Welling, C.~Cortes, N.~D. Lawrence, K.~Q. Weinberger
  (Eds.), Advances in Neural Information Processing Systems 27: Annual
  Conference on Neural Information Processing Systems 2014, December 8-13 2014,
  Montreal, Quebec, Canada, 2014, pp. 2672--2680.

\bibitem{Zhang2019}
H.~Zhang, H.~Chen, Z.~Song, D.~Boning, I.~S. Dhillon, C.-J. Hsieh, The
  limitations of adversarial training and the blind-spot attack (Jan. 2019).
\newblock \href {http://arxiv.org/abs/1901.04684} {\path{arXiv:1901.04684}}.

\bibitem{Uesato2018}
J.~Uesato, B.~O'Donoghue, P.~Kohli, A.~van~den Oord, Adversarial risk and the
  dangers of evaluating against weak attacks, in: J.~Dy, A.~Krause (Eds.),
  Proceedings of the 35th International Conference on Machine Learning, Vol.~80
  of Proceedings of Machine Learning Research, PMLR, 2018, pp. 5025--5034.

\bibitem{xue2022a}
S.~Xue, R.~Guo, K.~P. Bohn, J.~Matzke, M.~Viscione, I.~Alberts, H.~Meng,
  C.~Sun, M.~Zhang, M.~Zhang, R.~Sznitman, G.~El~Fakhri, A.~Rominger, B.~Li,
  K.~Shi, A cross-scanner and cross-tracer deep learning method for the
  recovery of standard-dose imaging quality from low-dose pet, Eur J Nucl Med
  Mol Imaging 49~(6) (2022) 1843--1856.
\newblock \href {https://doi.org/10.1007/s00259-021-05644-1}
  {\path{doi:10.1007/s00259-021-05644-1}}.

\bibitem{Reyes2020}
M.~Reyes, R.~Meier, S.~Pereira, C.~A. Silva, F.-M. Dahlweid, H.~v.
  Tengg-Kobligk, R.~M. Summers, R.~Wiest, On the interpretability of artificial
  intelligence in radiology: Challenges and opportunities, Radiology:
  Artificial Intelligence 2~(3) (2020) e190043.
\newblock \href {https://doi.org/10.1148/ryai.2020190043}
  {\path{doi:10.1148/ryai.2020190043}}.

\bibitem{Kundu2021}
S.~Kundu, Ai in medicine must be explainable, Nature Medicine 27~(8) (2021)
  1328--1328.

\bibitem{Toja2021}
E.~Tjoa, C.~Guan, A survey on explainable artificial intelligence (xai): Toward
  medical xai, IEEE Transactions on Neural Networks and Learning Systems
  32~(11) (2021) 4793--4813.
\newblock \href {https://doi.org/10.1109/TNNLS.2020.3027314}
  {\path{doi:10.1109/TNNLS.2020.3027314}}.

\bibitem{Ghassemi2021}
M.~Ghassemi, L.~Oakden-Rayner, A.~L. Beam, The false hope of current approaches
  to explainable artificial intelligence in health care, The Lancet Digital
  Health 3~(11) (2021) e745--e750.
\newblock \href {https://doi.org/10.1016/S2589-7500(21)00208-9}
  {\path{doi:10.1016/S2589-7500(21)00208-9}}.

\bibitem{eclair}
P.~Omoumi, A.~Ducarouge, A.~Tournier, H.~Harvey, C.~E. Kahn, F.~Louvet-de
  Verch{\`e}re, D.~Pinto Dos~Santos, T.~Kober, J.~Richiardi, {To buy or not to
  buy - evaluating commercial AI solutions in radiology (the ECLAIR
  guidelines)}, European radiology 31~(6) (2021) 3786--3796.

\bibitem{Buchgeister2021}
M.~Buchgeister, D.~Hummel, Risikoanalyse in der strahlentherapie: Muss es die
  fmea-methode mit rpz sein?, Zeitschrift für Medizinische Physik 31~(4)
  (2021) 343--345.
\newblock \href {https://doi.org/https://doi.org/10.1016/j.zemedi.2021.09.002}
  {\path{doi:https://doi.org/10.1016/j.zemedi.2021.09.002}}.

\bibitem{molnar2022}
C.~Molnar,
  \href{https://christophm.github.io/interpretable-ml-book}{{Interpretable
  Machine Learning}}, 2nd Edition, 2022.
\newline\urlprefix\url{https://christophm.github.io/interpretable-ml-book}

\bibitem{Arun2021}
N.~Arun, N.~Gaw, P.~Singh, K.~Chang, M.~Aggarwal, B.~Chen, K.~Hoebel, S.~Gupta,
  J.~Patel, M.~Gidwani, J.~Adebayo, M.~D. Li, J.~Kalpathy-Cramer, {Assessing
  the Trustworthiness of Saliency Maps for Localizing Abnormalities in Medical
  Imaging}, Radiology: Artificial Intelligence 3~(6) (2021) e200267.
\newblock \href {https://doi.org/10.1148/ryai.2021200267}
  {\path{doi:10.1148/ryai.2021200267}}.

\end{thebibliography}

\end{document}